# Gravitational-wave Limits from Pulsar Timing Constrain Supermassive Black Hole Evolution

**Authors:** R. M. Shannon[1,*,†], V. Ravi[2,1,*,†], W. A. Coles[3], G. Hobbs[1], M. J. Keith[1], R. N. Manchester[1], J. S. B. Wyithe[2], M. Bailes[4], N. D. R. Bhat[4,5], S. Burke-Spolaor[6], J. Khoo[1,7], Y. Levin[8], S. Osłowski[4], J. M. Sarkissian[9], W. van Straten[4], J. P. W. Verbiest[10], J-B. Wang[1,11]

**Affiliations:**

1. CSIRO Astronomy and Space Science, Australia Telescope National Facility, PO Box 76, Epping, NSW 1710, Australia.
2. School of Physics, University of Melbourne, Parkville, VIC 3010, Australia.
3. Department of Electrical and Computer Engineering, University of California at San Diego, La Jolla, CA 92093, USA.
4. Centre for Astrophysics and Supercomputing, Swinburne University of Technology, PO Box 218, Hawthorn, VIC 3122, Australia.
5. International Centre for Radio Astronomy Research, Curtin University, Bentley, WA 6102, Australia.
6. Jet Propulsion Laboratory, California Institute of Technology, 4800 Oak Grove Dr, Pasadena, CA 91109-8099, USA.
7. CSIRO Advanced Scientific Computing, Information Management and Technology, Private Bag 33, Clayton, VIC 3169, Australia.
8. School of Physics, Monash University, PO Box 27, VIC 3800, Australia.
9. CSIRO Astronomy and Space Science, Parkes Observatory, PO Box 276, Parkes, NSW 2870, Australia.
10. Max-Planck-Institut für Radioastronomie, Auf dem Hügel 69, 53121 Bonn, Germany.
11. Xinjiang Astronomical Observatory, CAS, 150 Science 1-Street, Urumqi, Xinjiang 830011, China.

\* Correspondence to: ryan.shannon@csiro.au, v.ravi@pgrad.unimelb.edu.au

† These authors contributed equally to the work.

**Abstract:**

The formation and growth processes of supermassive black holes (SMBHs) are not well constrained. SMBH population models, however, provide specific predictions for the properties of the gravitational-wave background (GWB) from binary SMBHs in merging galaxies throughout the Universe. Using observations from the Parkes Pulsar Timing Array, we constrain the fractional GWB energy density with 95% confidence to be $\Omega_{GW}(H_0/73 \text{ km s}^{-1} \text{ Mpc}^{-1})^2 < 1.3 \times 10^{-9}$ at a frequency of 2.8 nHz, which is approximately a factor of six more stringent than previous limits. We compare our limit to models of the SMBH population and find inconsistencies at confidence levels between 46% and 91%. For example, the standard galaxy formation model implemented in the Millennium simulations is inconsistent with our limit with 50% probability.

**Main Text:**

Supermassive black holes (SMBHs), with masses between $10^6$ and $10^{11}$ solar masses, are observed to exist at the centers of all massive galaxies in the nearby Universe, and to have masses that scale closely with properties of their hosts (*1, 2*). Together, these phenomena suggest that the growth processes of SMBHs and of their host galaxies are connected. Galaxies, and groups of galaxies, are embedded in even larger dark matter halos, which form and evolve through the hierarchical merging of smaller dark matter halos and galaxies (*3, 4*). Galaxy mergers are expected to result in binary SMBHs (*5, 6*), which, while notoriously difficult to observe via electromagnetic signatures, are expected to be the strongest sources of gravitational waves in the Universe (*7*). The universality of galaxy mergers implies the existence of a gravitational-wave background (GWB) from binary SMBHs (*8, 9*).

The GWB is manifested as a red-noise process in pulse arrival time measurements from pulsars (*10*). Pulsar timing array groups search for evidence of the GWB in radio-frequency observations of millisecond pulsars, which have rivaled the stability of the best clocks on Earth over timescales of tens of years (*11*). The GWB is commonly parameterized by its wave amplitude spectrum $h_c(f) = A(f/f_{yr})^{-2/3}$, where $f$ is the received gravitational-wave frequency, $f_{yr}$ is a reference frequency of one cycle per year, and $A$ is the characteristic amplitude that defines the strength of the GWB. The fraction of the critical energy density of the Universe, per logarithmic frequency interval, of the GWB is $\Omega_{GW}(f) = (2\pi^2/3 H_0^2) A^2 f_{yr}^2 (f/f_{yr})^{2/3}$ (*10*), where $H_0$ is the Hubble constant, which we assume to be 73 km s$^{-1}$ Mpc$^{-1}$. Recent observations by two separate pulsar timing array groups have been analyzed to find $A < 6 \times 10^{-15}$ (*12*) and $< 7 \times 10^{-15}$ (*13*) with 95% confidence.

We have been monitoring pulse arrival times from twenty millisecond pulsars with the 64-metre Parkes Telescope as part of the Parkes Pulsar Timing Array (PPTA) project (*14*) and previous observing programs (*15*). We extended the timing baseline of this data set by including publicly available observations from the Arecibo observatory (*16*). A detection of the GWB relies on measuring correlations between residual pulse arrival times for multiple pulsars with different angular separations on the sky. Within the PPTA timing programme, there are presently too few pulsars with sufficient timing precision and data span to make an unambiguous detection of the GWB feasible (*17*). We instead constrain the GWB amplitude using observations of six pulsars with the lowest noise levels over the longest observing spans (*18*) (Fig. 1).

Our limit on the strength of the GWB was computed in two stages (*19*). For each pulsar $j$, we first estimated the power spectral density, $P_j(f_i)$, of the residual pulse arrival times, after a fit for a pulsar model (*20*), at frequencies $f_i$ that are harmonics of $1/T_{obs}$, where $T_{obs}$ is the observing span for the pulsar. A prewhitening method (*21*) was used in the spectral estimation to eliminate spectral leakage and to provide nearly independent spectral estimates even if red-noise signals, such as those expected from the GWB, are present. We form a detection statistic (DS) from the power spectra:

$$\hat{A}^2 = \sum_{ij} [P_j(f_i) g_j(f_i)/M_j(f_i)^2]/\sum_{ij} [g_j(f_i)/M_j(f_i)]^2, \qquad (1)$$

where $g_j(f_i)$ is the shape of the power spectrum induced by the GWB, and $M_j(f_i)$ is a model of the observed spectrum (Fig. 1). The DS, $\hat{A}^2$, combines individual spectral estimates, $P_j(f_i)$, to form a conservative estimate of the square of the characteristic amplitude of the GWB, $A^2$. If the spectral models are correct, a DS of the form in Eq. (1) provides an estimate of $A^2$ with a maximal signal to noise ratio. To set a limit on $A$, we compared the observed value of the DS with distributions of the DS derived from simulated data sets, which include white noise consistent with the observations and a GWB of strength $A_{sim}$. Many trial simulations were conducted at a given $A_{sim}$ to account for the stochasticity of the GWB. The 95% confidence limit on the GWB amplitude, $A_{95}$, is the value of $A_{sim}$ at which only 5% of the $\hat{A}^2$ trials are lower than the observed $\hat{A}^2$.

We simulated both Gaussian (*10*) and non-Gaussian (*9*) GWB-induced residual pulse arrival times. Although previous pulsar timing array limits on the strength of the GWB (*12, 13*) were derived assuming Gaussian statistics, a non-Gaussian background, dominated by fewer binary SMBHs, is predicted from some models of the binary SMBH population (*8, 9*).

We verified the efficacy of the algorithm by correctly bounding the GWB strength in synthetic data sets, including those in the International Pulsar Timing Array Data Challenge and other mock data sets that contained features of the observations such as inhomogeneous observing cadence, highly heteroscedastic pulse arrival times, and red noise (*22*). When applied to the PPTA data set, and assuming a Gaussian GWB, we find that $\Omega_{GW}(f_{PPTA})(H_0/73$ km s$^{-1}$ Mpc$^{-1}$ )$^2 < 1.3 \times 10^{-9}$ with 95% confidence at a gravitational-wave frequency of $f_{PPTA}$ = 2.8 nHz (*23*). This is equivalent to $A_{95} = 2.4 \times 10^{-15}$. Compared to the power spectra, $P_j$, of the measured residual pulse arrival times, the mean power spectra of 200 simulated realizations with $A_{sim} = A_{95}$ (displayed in Fig. 1 as green lines) show, as expected, excess power at the lowest frequencies. For a non-Gaussian GWB, we find $\Omega_{GW}(f_{PPTA})(H_0/73$ km s$^{-1}$ Mpc$^{-1}$ )$^2 < 1.6 \times 10^{-9}$ with 95% confidence, corresponding to $A_{95} = 2.7 \times 10^{-15}$.

The PPTA bound on the GWB enables direct tests of models for galaxy and SMBH formation that specify the population of binary SMBHs in the Universe. We compared the probability, Pr($\Omega_{GW}$), that a GWB of energy density $\Omega_{GW}(f_{PPTA})$ exists given the PPTA observations with four predictions for the GWB from binary SMBHs, expressed as the probability density function of $\Omega_{GW}(f_{PPTA})$, $\rho_M(\Omega_{GW})$ ( *24*) (Fig. 2). All four predictions account for the most recent SMBH mass and galaxy bulge mass measurements, and include the assumption that all binary SMBHs that contribute to the GWB are in circular orbits and not interacting with their environments.

First, a model that assumes a scenario in which all evolution in the galaxy stellar mass function and in the SMBH mass function is merger-driven at redshifts $z < 1$ ( *25*) predicts a Gaussian GWB that is ruled out at the 91% confidence level. However, the assumption of purely merger-driven evolution leads to the largest possible GWB amplitude given observational data.

A synthesis of possible combinations of current observational estimates of the galaxy merger rate and SMBH-galaxy scaling relations results in a large range of possible GWB amplitudes (*26*). PPTA observations exclude 46% of this set of GWB amplitudes, assuming a Gaussian GWB.

As a specific example for how pulsar timing array observations can impact models of SMBH formation and growth, we calculated the level of $\Omega_{GW}(f_{PPTA})$ ( *24*) expected from a semi-analytic galaxy formation model (*4*) implemented within the Millennium (*27*) and Millennium-II (*28*) dark matter simulations. This model, where SMBHs are seeded in every galaxy merger remnant at early times and grow primarily by gas accretion triggered by galaxy mergers, represents the standard paradigm of galaxy and SMBH formation and evolution. The model accurately reproduces the luminosity function of quasars at $z < 1$ corresponding to the epoch predicted to dominate the GWB (*8, 25, 26*). The range of predictions for $\Omega_{GW}(f_{PPTA})$ results from the finite observational sample of measured SMBH and bulge mass pairs (*2*), which is used to tune the model, but neither accounts for uncertainties in the observed galaxy stellar mass function (*4*) nor for uncertainties in the nature of the relations between SMBH masses and bulge masses (*2*). Assuming a non-Gaussian GWB, the probability of this prediction for $\rho_M(\Omega_{GW})$ being inconsistent with the PPTA data is 49%.

A complementary prediction for the strength of the GWB comes from an independent model for SMBH growth at late times (*29*). This model examines the growth mechanisms of SMBHs in cluster and void environments through mergers and gas accretion. The model is inconsistent with the PPTA data at the 61% confidence level.

The PPTA constraints on the GWB show that pulsar timing array observations have reached a sufficient level of sensitivity to test models for the binary SMBH population. The highest galaxy merger rate that is consistent with the observed evolution in the galaxy stellar mass function (*25*) is inconsistent with our limit. We exclude 46% of the parameter space of a model that surveys empirical uncertainties in the growth and merger of galaxies and black holes (*26*), and our results therefore reduce these uncertainties. Although the PPTA limit only excludes 49% and 61% of realizations of the GWB from two galaxy and SMBH evolution models, these models are open to refinement.  For example, these models do not include SMBH formation mechanisms consistent with high-redshift quasar observations (*30*), nor do they reproduce the observed larger scatter and possibly higher normalization in SMBH-galaxy scaling relations for the most massive SMBHs (*1, 2*).  Other physical effects will also be built into the next generation of GWB models. For example, recent numerical simulations of massive galaxy mergers predict binary SMBHs with eccentricities ranging between 0.1 ( *31*) and 0.9 ( *32*). If binaries radiating gravitational waves at frequencies relevant to pulsar timing arrays are significantly eccentric or predominantly evolving under environmental interactions (*33*), the spectral shape of $\Omega_{GW}(f)$ may differ from current predictions (*34*).

**Acknowledgements.** We thank all of the observers, engineers, and Parkes observatory staff members who have assisted with the observations reported in this paper.   We thank N. McConnell for providing and confirming some dynamical SMBH and bulge mass measurements, S. Mutch for discussions on the Millennium-based model, and X-J. Zhu for comments on the manuscript. The Parkes radio telescope is part of the Australia Telescope National Facility which is funded by the Commonwealth of Australia for operation as a National Facility managed by CSIRO. The Millennium and Millennium-II Simulation databases used in this paper and the web application providing online access to them were constructed as part of the activities of the German Astrophysical Virtual Observatory. The PPTA project was initiated with support from RNM's Australian Research Council (ARC) Federation Fellowship (FF0348478) and from the CSIRO under that Fellowship program. The PPTA project has also received support from ARC Discovery Project grant DP0985272. VR is a recipient of a John Stocker Postgraduate Scholarship from the Science and Industry Endowment Fund, GH is the recipient of an ARC QEII Fellowship (DP0878388), and JSBW acknowledges an Australian Research Council Laureate Fellowship. Part of this research was carried out at the Jet Propulsion Laboratory, California Institute of Technology, under a contract with the National Aeronautics and Space Administration. JPWV acknowledges the financial support by the European Research Council for the ERC Starting Grant *Beacon* under contract no. 279202. The authors declare no conflicts of interest. Data used in this analysis can be accessed via the Australia National Data Service (http://www.ands.org.au).

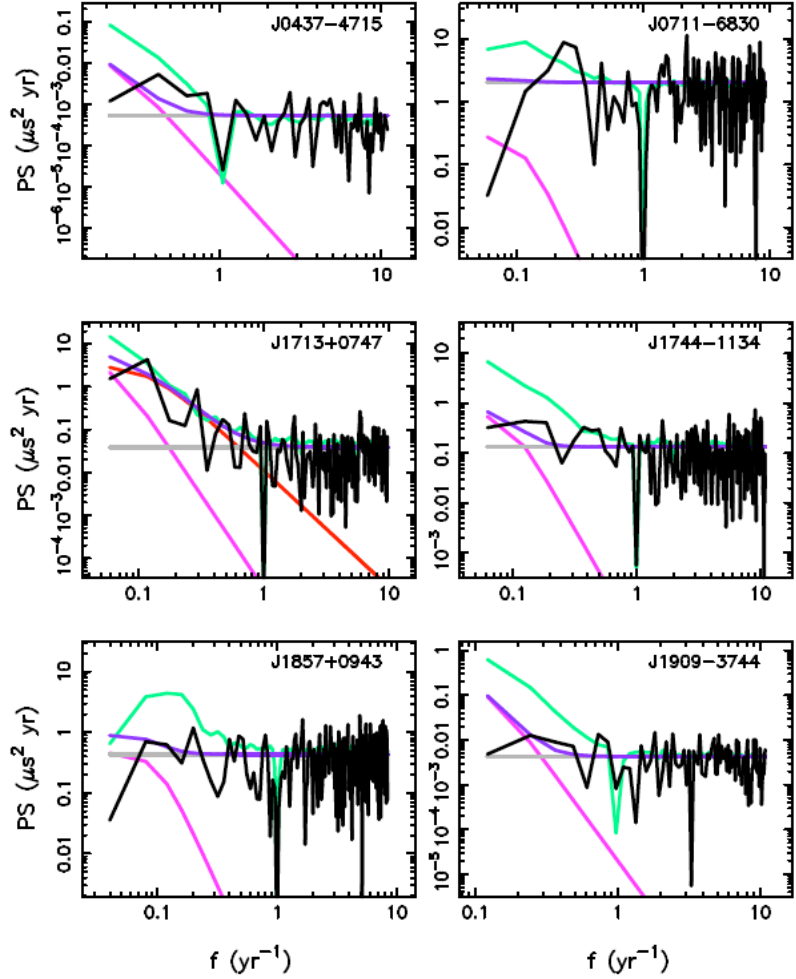

**Fig. 1:** Observed power spectra of the residual post-fit arrival times, and models of these spectra, for the PPTA pulsars used to set the limit on the GWB amplitude. The observed power spectra (PS), $P_j$, for the pulsars are shown as black lines, along with the models of the PS, $M_j = W_j + G_j + R_j$ (shown as purple lines). The models contain a white component ($W_j$, gray lines), a common GWB component $G_j$ (pink lines), and, for PSR J1713+0747, an additional red-noise term $R_j$ (red line). The PS models were only used for the determination of the weights in the calculation of the detection statistic. The green curves show what the PS would look like (on average) in the presence of a Gaussian GWB with amplitude $2.4 \times 10^{-15}$.

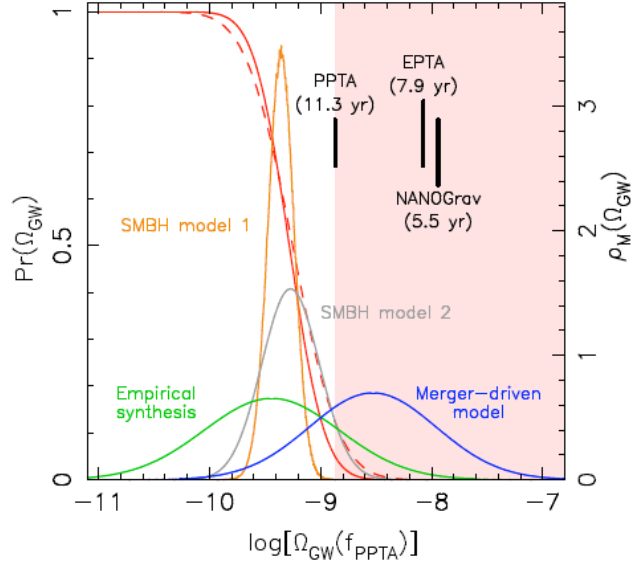

**Fig. 2**: Comparison between the PPTA constraints on $\Omega_{GW}(f_{PPTA})$ and various model predictions (*24*). The probabilities, $Pr(\Omega_{GW})$, given the PPTA data, that a GWB relative energy density $\Omega_{GW}(f_{PPTA})$ exists, assuming Gaussian (*10*) and non-Gaussian (*9*) GWB statistics, are shown as red solid and dashed lines respectively. The pink shaded area represents the values of $\Omega_{GW}(f_{PPTA})$ ruled out with greater than 95% confidence assuming a Gaussian GWB. The labeled curves represent the probability density functions, $\rho_M(\Omega_{GW})$, for $\Omega_{GW}(f_{PPTA})$ predicted by a synthesis of empirical models (*26*) (green), assuming merger-driven galaxy evolution at redshifts $z<1$ (*25*) (blue), from the semi-analytic galaxy formation model (SMBH model 1, orange) that we discuss in the text, and from a second distinct model (*29*) for SMBH growth (SMBH model 2, gray). When integrated over $\Omega_{GW}$, the product of $Pr(\Omega_{GW})$ and $\rho_M(\Omega_{GW})$ gives the probability of the model being consistent with the data. The vertical bars indicate the 95% confidence upper limits on $\Omega_{GW}(f_{PPTA})$ assuming a Gaussian GWB from the PPTA, and recently published limits from the European Pulsar Timing Array (EPTA, *12*) and the North American Nanohertz Observatory for Gravitational Waves (NANOGrav, *13*) scaled to $f_{PPTA}$. The times next to the limits correspond to the reciprocal of the frequency of maximum sensitivity and are approximately the observing span of the data sets (*12*, *13*, *23*).

# Supplementary Materials for

## Gravitational-wave Limits from Pulsar Timing Constrain Supermassive Black Hole Evolution

R. M. Shannon, V. Ravi, W. A. Coles, G. Hobbs, M. J. Keith, R. N. Manchester, J. S. B. Wyithe, M. Bailes, N. D. R. Bhat, S. Burke-Spolaor, J. Khoo, Y. Levin, S. Osłowski, J. M. Sarkissian, W. van Straten, J. P. W. Verbiest, J-B. Wang
Correspondence to: ryan.shannon@csiro.au, v.ravi@postgrad.unimelb.edu.au

**This PDF file includes:**

> Supplementary Text
> Fig. S1
> Tables S1 and S2
> References

Supplementary Text

These supplementary materials accompany the *Science Report* by R. Shannon et al. on the Parkes Pulsar Timing Array (PPTA) limit on the amplitude of the gravitational-wave background (GWB) from supermassive black hole (SMBH) binaries. We present details of the pulsar timing data sets used in this study in Section S1, and a description of the algorithm we constructed to estimate the upper bound on the GWB is given in Section S2. We present the calculation of the gravitational-wave frequency, $f_{\rm PPTA}$, of the upper limit in Section S3. We discuss inconsistencies in definitions of the gravitational wave strain from binary SMBHs in Section S4. Summaries of the existing empirical models for the GWB that we consider in the main text (*25, 26*) are given in Section S5. A description of the semi-analytic modeling of the GWB based on the Millennium simulations is given in Section S6, as well as an outline of the other physical model we consider (*29*). In Section S7, we outline the calculation of the fractions of realizations of the GWB from different models that are inconsistent with the PPTA data.

### S1: Data sets used to limit the gravitational wave background

The technique of *pulsar timing* relies on the precise measurement of the pulse times of arrival (ToAs) from a given pulsar. The differences between the measured ToAs and the predictions from a best-fit model for the ToAs are referred to as the *timing residuals*. Significant non-zero residuals indicate a poorly determined model or the presence of one or more processes affecting the ToAs not in the timing model. Gravitational waves (GWs) passing between the pulsar and Earth are one such process. A GWB emitted from a population of binary SMBHs is expected to lead to timing residual time-series that exhibit red-noise power spectra (*10*). We place an upper limit on the amplitude of any such GWB by measuring the power spectra of timing residuals from long-term observations of six pulsars, and by comparing the measured power spectra with those expected from a GWB of specified amplitude. Details of this algorithm are presented in Section S2.

We have used observations from the initial (DR1) and extended (DR1E) data releases from the PPTA project (*14*). The DR1 data set comprises ToA data spanning six years that have been optimally corrected for fluctuations in interstellar dispersion, DM(*t*) (*35*), using observations that cover a wide range of observing frequencies. The DR1E data set includes the DR1 data supplemented by archival observations that significantly increase the data span. However, most of these earlier data were obtained without sufficient frequency coverage to enable corrections for DM(*t*). None of the DR1E data sets were published with DM(*t*) corrections to earlier data.

In total, the data sets include observations of 20 pulsars; however, our constraint on the strength of the GWB is dominated by the six pulsars that have the lowest noise levels. We get consistent results if we include the additional pulsars in the algorithm, but for simplicity do not present that analysis here. The six pulsars we use are listed in Table S1, and the measured timing residuals for each pulsar are shown in Fig. S1. Details of the observations are provided elsewhere (*14*).

Analyses of the power spectra of the timing residuals (see Section S2 for descriptions on how the power spectra are formed) indicate that, for all of the six pulsars used here, the residuals exhibit white noise in excess to that expected from the corresponding arrival time estimation errors alone. This excess is not unexpected and can have a number of origins, including instrumental effects, imperfect calibration (*36*), propagation effects in the interstellar medium (*37*), and intrinsic pulse shape variations (*38, 39*). To account for this excess white noise, we have added, in quadrature, to the measured estimation errors an additional error term, which is often referred to as EQUAD. We note that this procedure does not alter the ToAs in any way: the procedure simply corrects the ToA uncertainties and results in unbiased estimation errors for the timing model parameters.

The required value of EQUAD for each pulsar was determined by comparing power spectra formed from simulated data sets to the actual power spectra obtained from the observations. The simulated data sets had the same cadence as the observations, but contained ToA errors that were the quadrature sums of the measured ToA errors and trial values of EQUAD. We adjusted the EQUAD levels until the white noise floor (set to be at frequencies $f > 2 \text{ yr}^{-1}$) in the average of 200 simulated power spectra matched the observed power spectra. In Table S1, we list the values for the EQUADs that were determined for each of the six pulsars used in the subsequent analysis.

Red-noise processes, in addition to being caused by gravitational radiation, may be associated with intrinsic rotational instabilities in the pulsar (*40*), or unmitigated effects of the propagation of radio waves through the interstellar medium (*35, 37*). Because of these effects, not all pulsars are suitable for pulsar timing array observations. Of the data sets used in this analysis, only one pulsar (PSR J1713+0747, noted below) shows red-noise levels inconsistent in shape and amplitude with a GWB.

For all of the pulsars, we have used the *best-band* ToAs from the DR1 data set. The best band is the observing band that provides the highest timing precision ToAs. We list these bands in the last column of Table S1. The timing analyses presented here and published as part of DR1 and DR1E differ in a few important ways. These differences are listed below:

*PSR J0437−4715:* To produce the limit we have used the shorter DR1 data set, which has been corrected for DM(*t*). In the DR1E data a large, red signal is present. Based on measurements of DM(*t*) in the DR1 data set, the red signal in the DR1E data set has an amplitude and spectral shape consistent with being induced by the DM(*t*) process observed in DR1. As a result, the shorter DR1 time series is more sensitive to the GWB than the longer, partially corrected DR1E time series; we therefore omit the DR1E data.

*PSR J0711−6830:* For this pulsar, we have used the DR1E best-band data set. Despite relatively large ToA uncertainties (typically approximately 1 μs) this pulsar is sensitive to the GWB because of the long timing baseline and the absence of red noise.

*PSR J1713+0747:* For this pulsar, we have used the DR1E data set. The power spectrum of the timing residuals for this pulsar shows evidence for a red-noise process (see Fig. 1). However, the red-noise spectrum has an amplitude and a slope that are inconsistent with the spectrum expected from a GWB.

*PSR J1744−1134:* We have modified the DR1E timing model by removing the DM($t$) correction of the DR1 observations. In the DR1 observations the pulsar shows a constant slope in DM($t$), which is common for many pulsars (*35*). For single-wavelength data, like those used here, a linear DM variation induces a linear trend in ToAs that is absorbed in the fit for pulsar spin frequency. In contrast, if the data are only partially corrected, as is the case for the DR1E dataset, then the effects of DM($t$) are not completely removed by the fit. We therefore did not correct for DM variations in this data set.

*PSR J1857+0943:* We used archival observations from the Arecibo telescope (*16*) in addition to the DR1 observations to extend the timing baseline of PSR J1857+0943 from 7 to 25 years. There is a large gap between the end of the Arecibo data set in 1994 and the commencement of the Parkes observations in 2003. The large gap, accompanied by an unknown instrumental offset, distorts the shape of the timing residual power spectrum induced by the GWB. However, as discussed in the next section, this distortion can be modeled.

*PSR J1909−3744:* We have used the DR1E data set, but with two important changes. First, we have utilized additional archival low-frequency 50 cm observations from the Parkes Telescope that extend to the beginning of the data set. These extra observations allow us to measure and correct for DM($t$) over the earlier part of the DR1E span. Second, we identified a 1 μs offset in 10 cm observations with one of the receiving backend instruments (PDFB1) that is not present in contemporaneous 20 cm observations with another system (CPSR2). We attribute this offset to an unmodeled instrumental delay in early 10 cm PDFB1 data. We therefore use the contemporaneous CPSR2 observations in the 20 cm band prior to MJD 53819.

### S2: Algorithm for bounding the strength of the GWB

Here we describe the algorithm that was used to place an upper bound on the strength of any isotropic, stochastic GWB at GW frequencies accessible to pulsar timing arrays. The algorithm relies on power spectral analyses of the pulsar timing residuals.

#### S2.1 A test statistic for estimating the strength of the GWB

For each pulsar, $j$, we first calculate the post-fit residual ToAs. From these we form the power spectra, $P_j(f_i)$, of the residuals at frequencies $f_i$. A prewhitening method (*21*) is used in the spectral estimation to eliminate spectral leakage and provide nearly independent spectral estimates at the harmonics of $f = 1/T_{obs}$, where $T_{obs}$ is the observing span. This procedure works even if red-noise signals, such as those expected from the GWB, are present.

The power spectra for the six pulsars used in this paper are displayed in Fig. 1. The power spectral estimates, $P_j(f_i)$, provide nearly-independent estimates, $\hat{A}^2_{ij}$, of $A^2$ following the relationship

$$\hat{A}^2_{ij} = P_j(f_i)/g_j(f_i) \, , \quad (1)$$

where $g_j(f_i)$ is a function that describes the shape of the GWB-induced residual power spectral density.

The power spectral density for the GWB is $G_j(f_i) = A^2 g_j(f_i)$ for a background with characteristic strain $A$. If we consider the GWB-induced ToA variations, rather than the post-fit residuals, $g_j(f_i) \propto f^{-13/3}$. However, the shape of the GWB-induced power spectrum in the residuals is distorted by a number of processes. The calculation of $g_j(f_i)$ is discussed in supplementary section S2.2.

If the only source of non-zero residuals was the GWB, then the estimators $\hat{A}^2_{ij}$ would be unbiased and would be maximum likelihood estimators of $A^2$. In the presence of any extra red or white noise the bias is non-negative, because the contributions of any white or red-noise processes to the spectral estimates, $P_j(f_i)$, are non-negative. This makes the individual estimators, $\hat{A}^2_{ij}$, conservative for our purpose. We derive an overall estimator of $A^2$, which we refer to as the detection statistic, $\hat{A}^2$, from the weighted average of these $\hat{A}^2_{ij}$ estimates,

$$\hat{A}^2 = \sum_{ij} \hat{A}^2_{ij} K_{ij} / \sum_{ij} K_{ij}, \quad (2)$$

where the weights $K_{ij}$ are chosen to maximize the signal to noise ratio in the estimation of $A$ from individual estimates $\hat{A}^2_{ij}$.

We ignore the non-GWB noise processes in determining the estimates, $\hat{A}^2_{ij}$, but include them when determining the weights, $K_{ij}$. This is necessary because the noise levels for some pulsars are much higher than for others. If these extra noise processes are ignored when forming the weights, serious degradation in the variance of the detection statistic occurs. This would result in a spuriously high bound on the amplitude of the GWB. Because each spectral estimate represents a $\chi^2$ random variable with two degrees of freedom, the variance of any spectral estimate is equal to the square of its mean, so

$$K_{ij} = [g_j(f_i) / M_j(f_i)]^2, \quad (3)$$

where $M_j(f_i) = G_j(f_i) + W_j + R_j(f_i)$ is a smoothed model of the power spectrum that contains the GWB, $G_j(f_i)$, of model strength $A_M$, white noise $W_j$, and, if present, red noise $R_j(f_i)$. To model the power spectra, we conduct a joint fit of all of the power spectra using a non-linear fitting algorithm in order to estimate the common parameter $A_M$. When applied to the PPTA data sets described in Section S1, we find $A_M = 1.2 \times 10^{-15}$.

It is necessary to estimate $M_j(f_i)$ from the observations; this process is inherently uncertain. However, by restricting this uncertainty to the weights we ensure that the upper bound cannot be biased downward by an error in estimating $R_j(f_i)$. At worst, it can increase the variance of the detection statistic and thus raise the upper limit. Combining Supplementary Eqs. (2) and (3), we use as a detection statistic

$$\hat{A}^2 = \sum_{ij} [P_j(f_i) \, g_j(f_i)/M^2_j(f_i)] \, / \sum_{ij} [g_j(f_i) /M_j(f_i)]^2 \, . \quad (4)$$

A derivation of a similar test statistic can be found in section 5.3.2 of Ref. (*41*).

### S2.2 Calculating the shape of GWB power spectrum

The shape of the GWB-induced power spectrum in the residuals is distorted by the fitting procedure, uneven sampling and heteroscedasticity of the data. These effects cause the GWB-induced power in the lowest frequency bins to be attenuated by a factor generally less than five. The shape of the distortion for each pulsar can be measured through a series of simulations of pulse ToAs (with an identical cadence to the actual observations) containing both the measured white noise and a GWB of strength $A_M = 1.2\times10^{-15}$, chosen to be the strength of the modeled GWB signal, discussed below.

We calculated the power spectra of the simulated ToAs using our prewhitening algorithm (*21*). The shape, $g_j(f_i)$, was calculated from the average of power spectra from 200 realizations. A subsequent set of simulations was used to confirm that this modeling does not affect our bounding technique.

### S2.3 Setting a limit on the strength of the GWB

To place an upper limit on the GWB strength, we compared the observed detection statistic to distributions of detection statistics obtained from a series of simulations that each contains a GWB of strength $A_{\text{sim}}$ and white noise. For the simulations, we used an identical observing cadence to the observations and generated white noise at levels consistent with observations. We did not simulate the red noise present in the PSR J1713+0747 data set. This choice only conservatively biases our bound. These simulated data sets were then processed using the same fitting algorithm that was applied to the actual data sets. Using the same weighting function $K_{ij}$ as for the observations, we calculate the distribution of simulated detection statistics.

The strength of the GWB was adjusted until a level $A_{\text{sim}} = A_{95}$ was found such that in 95% of the simulations, the detection statistic exceeded the observed detection statistic. From the observations, we find $\hat{A} = 1.6\times10^{-15}$ and $A_{95} = 2.4\times10^{-15}$, both of which are consistent with $A_M$. The value of the observed detection statistic is interesting, but one must bear in mind that any amount of non-GWB red noise may be present, so our results are consistent with no GWB being present.

We tested the algorithm using a number of mock data sets, including the six data sets that were part of the first International Pulsar Timing Array (IPTA) Data Challenge. All of the data sets comprised simulated ToAs for 38 pulsars observed as part of the IPTA (*42*), spanning five years. The data sets differed in observing cadence, strength of the GWB with spectral index $\alpha = -2/3$, and levels of additional noise in the ToAs. Three *open* data sets contained background amplitudes published at the time of release. In contrast, for the three *closed* data sets, the background levels were announced approximately six months after the release. In Table S2, we show both the levels of the GWB contained in the simulated data sets (*A*) and the 95% confidence limits we reach ($A_{95}$). In every case our limit is above the injected value, which is consistent with our algorithm producing a 95% confidence limit.

We also tested the algorithm using data sets that better matched our observations. The simulated data sets had identical cadence to the observed data sets, white noise consistent with observations, a GWB of amplitude of $A_{\text{sim}} = 2.5\times10^{-15}$, and low levels of red noise. We simulated 100 data sets and found in 99 of the cases our $A_{95} > A_{\text{sim}}$ confirming that our algorithm is conservative.

In placing our limit, we simulated GWBs that followed both Gaussian (*10*) and non-Gaussian (*9*) statistics. Limits (*12, 13*) are typically placed assuming that the GWB-induced ToA variations are a Gaussian random process. For a Gaussian GWB, we find a 95% confidence limit of $A_{95}$ = $2.4 \times 10^{-15}$. However, recent modeling of the GWB arising from SMBH binaries suggests that significant departures from Gaussian statistics may occur (*8, 9*). The non-Gaussian background shows larger realization-to-realization variation in strength of the GW signal in the simulated ToAs, and hence results in a slightly higher limit. For the non-Gaussian background, we find $A_{95}$ = $2.7 \times 10^{-15}$.

In order to make a definitive detection of the background it would be necessary to search for the expected correlations between the residual ToAs of different pulsars (*17, 43*). However there are currently an insufficient number of pulsars sensitive to the GWB at the levels constrained by our upper limit to search for the angular correlation. It is worth noting that, as our result demonstrates, only a few well-timed pulsars are needed to set a constraining upper limit.

### S3: Calculating $f_{PPTA}$

As outlined in the main text, our limit on the strain spectrum of the GWB of the form $h_c(f) = A f^\alpha$ (where $\alpha = -2/3$ for the SMBH binary GWB) can be related to the energy density per logarithmic frequency interval $\Omega_{GW}(f_{PPTA})$ at a center frequency $f_{PPTA}$. To make the conversion, we calculate the reference GW frequency $f_{PPTA}$ using a technique (*13*) that has previously been applied to pulsar timing array data sets. Under the assumption that the limit is set in a narrow frequency range, the limiting strain amplitude, $A_{95}$, will be proportional to $(f_{PPTA})^\alpha$. This assumption is valid because the spectrum of the signal induced by the GWB in timing residual power spectra is steep, so the limit is set on gravitational radiation emitted close to $1/T_{obs}$. It is not exactly $1/T_{obs}$ for two reasons. Firstly, our data sets vary in total length, so a global $T_{obs}$ is poorly defined. Secondly, the fitting process removes some of the power in the lowest frequency bins shifting $f_{PPTA}$ to frequencies slightly greater than $1/T_{obs}$.

To find $f_{PPTA}$, we therefore calculated $A_{95}$ for backgrounds with six values of $\alpha$ ranging from $-0.5$ to $-1.17$. To these points, we fit the relationship $A_{95} = A_0 (f_{PPTA})^\alpha$ to find an unimportant scaling factor, $A_0$, and $f_{PPTA}$.

### S4: A consistent definition of the GWB amplitude

There are inconsistencies in published derivations of the amplitude of the GWB from binary SMBHs. Some analyses (*9, 25*) use an expression for the orientation and polarization-averaged strain from a binary SMBH which includes a factor of $(4/3)^{1/2}$, while others (*8, 26, 29*) do not include this factor. The inclusion of the factor increases the predicted strength of the GWB. In the case of ground-based gravitational wave detectors (*44*), the factor is used to account for the rotation of the Earth when estimating the signal to noise ratio; it should not be included when calculating a strain spectrum.

We confirmed the absence of this factor in a number of ways. First, it is possible to calculate the spectral density of the GWB, $S_h(f)$, using two independent methods, either by using Equation 4 of (*45*) or by expressing it directly in terms of the mean squared strain amplitude, $h_s^2$, for each binary, as

$$S_h(f) = \int_0^\infty dz N(z) \frac{4\pi d^2 V_C}{d\Omega dz} \frac{dz}{dt} \frac{dt}{df} h_s^2, \quad (5)$$

where $N(z)$ is the number of binaries per unit commoving volume per unit redshift, $4\pi d^2 V_c/(d\Omega\, dz)$ is the co-moving volume shell between redshifts $z$ and $dz$, and $t$ is observer time. This equation and Equation 4 of (*45*) are equivalent only if the factor of $(4/3)^{1/2}$ is absent in the expression for $h_s$. Next, $S_h(f)$ is related to the power spectral density of the GWB-induced ToA variations, $P(f)$, as (Ref. *10*)

$$P(f) = \frac{1}{12\pi^2} \frac{1}{f^2} S_h(f). \quad (6)$$

We confirmed this relation and the absence of the factor of $(4/3)^{1/2}$ in the specific case of a GWB from binary SMBHs by analytically calculating the variance of the ToA variations induced by an individual binary in terms of its chirp mass, orbital frequency and distance, with random orientation parameters. We also simulated the effects on mock pulsar timing datasets of populations of binaries with random orientation parameters in the TEMPO2 software package (*10*), and checked that the power spectra of the residuals, calculated using a similar prewhitening method to that used with the real data (*21*), were consistent with the above calculations.

In order to directly compare all of the predictions for the GWB, we re-scaled predictions that included this factor. Therefore the prediction for $\Omega_{\rm GW}(f_{\rm PPTA})$ from (*25*) was multiplied by a factor of 3/4.

### S5: Empirical models for the GWB

In the main text, we compare the PPTA constraints on $\Omega_{\rm GW}(f_{\rm PPTA})$ with the two most recent predictions for $\Omega_{\rm GW}(f_{\rm PPTA})$ directly based on observations. These are as follows:

1. *An empirical synthesis of models* (Ref. *26*): This work considered the predicted GWB amplitudes within the empirically constrained parameter space of galaxy and supermassive black hole co-evolution. The strength of the assumed Gaussian GWB is estimated using galaxy stellar mass functions, galaxy close-pair fractions, galaxy merger timescales, and SMBH-galaxy scaling relations. We restricted our analysis to the predicted range of GWB amplitudes given the most recent SMBH and bulge mass measurements, and assume a Gaussian GWB.

2. *A merger-driven galaxy evolution model* (Ref. *25*): This work predicts the GWB assuming that all evolution in the observed galaxy stellar mass function and the SMBH mass function at redshifts $z<1$ is caused by galaxy mergers alone. This naturally leads to the highest possible prediction for the GWB given empirical constraints (*26*), because it results in a maximal galaxy merger rate. In our analysis, we use a Gaussian GWB, as suggested in Ref. (*25*). We also assume that the prediction, made at GW frequencies greater than $f_{\rm PPTA}$, can extrapolated to $f_{\rm PPTA}$ by assuming $\Omega_{\rm GW}(f) \propto f^{2/3}$, which is expected for binary SMBHs in circular orbits evolving under GW emission alone. Finally, we scale this prediction for $\Omega_{\rm GW}(f_{\rm PPTA})$ by 3/4 as described above.

### S6: Physical models for the GWB
#### S6.1: A model for the GWB based on the Millennium simulations

Here, we assume the following cosmological parameters: a fractional matter density of $\Omega_M = 0.25$, a fractional dark energy density of $\Omega_\Lambda = 0.75$, and a Hubble constant of $H_0 = 73$ km s$^{-1}$ Mpc$^{-1}$. These values are consistent with those used in the Millennium simulations (*27, 28*). The Millennium simulations use slightly different cosmological parameters to those recently measured by the Wilkinson Microwave Anisotropy Probe (*46*). These differences have a negligible effect on galaxy merger rates for the redshifts $z < 3$ relevant to this work (*46*), and therefore do not affect our results.

We modeled the GWB from binary SMBHs using the semi-analytic galaxy formation model of Guo et al. (Ref. *4*, hereafter G11). The semi-analytic model describes the evolution of baryonic matter within the evolving cold dark matter distribution derived from the Millennium (*27*) and Millennium-II (*28*) simulations. SMBHs at the centers of galaxies in the model grow primarily through galaxy mergers. In a galaxy merger, the two central SMBHs first coalesce, after which the resulting SMBH accretes dynamically cold gas of mass

$$\Delta M_{BH} = f_{BH} \frac{M_{min}}{M_{max}} \left[ \frac{M_{cold}}{1 + \left(280\ km\ s^{-1}/V_{vir}\right)^2} \right], \quad (7)$$

where $M_{min}$ and $M_{maj}$ are, respectively, the baryonic masses of the secondary and primary merging galaxies; $M_{cold}$ is the cold gas mass of the merged galaxy, $V_{vir}$ is the virial velocity of the merged dark matter halo; and $f_{BH}$ is a free parameter of the model that sets the fraction of cold gas accreted in each merger. Predictions of the G11 model are consistent with many properties of galaxies across cosmic time. The predicted SMBH mass function at different redshifts, and hence the SMBH merger rate, is directly related to the galaxy stellar mass, luminosity and color distributions. The SMBH growth history predicted by the model was also used to produce a model bolometric quasar luminosity function that reasonably matches observations for redshifts $z < 3$ (Ref. *47*).

**S6.2 Including new SMBH and bulge-mass measurements in the model**

Central to the G11 model treatment of SMBHs is its ability to reproduce the relationship between galaxy bulges and SMBH masses observed in the local Universe. The $M_{BH} - M_{bulge}$ relation was recently re-calculated by McConnell & Ma (Ref. *2*, hereafter MM13) using SMBH and bulge-mass estimates for a larger sample of galaxies than used in previous studies. In a number of galaxies, existing measurements of $\Gamma = M_{BH}/M_{bulge}$ were revised to higher values, resulting in a normalization for the $M_{BH} - M_{bulge}$ relation that is a factor of 1.8 greater than previously calculated (*48*).

These new mass measurements require the alteration of two covariant quantities in the G11 model: the amounts of cold gas accreted by central SMBHs following each merger, parameterized by $f_{BH}$; and the quiescent accretion rate onto central SMBHs from hot gas halos, parameterized by $\kappa_{AGN}$ in the G11 model. The former parameter governs the dominant growth mechanism of SMBHs. The accretion of hot gas, governed by the latter parameter, is associated with the suppression of cooling of the hot gas, and hence a suppression of quiescent star formation. These parameters were set in an earlier iteration of the G11 model (*3*) using an older $M_{BH} - M_{bulge}$ relation (*48*) and the observed galaxy stellar mass function.

However, the large covariance between the two parameters (*49*) implies that the parameters can be tuned to the new $M_{BH} - M_{bulge}$ relation without affecting the self-consistency of the model. We characterize the updated SMBH-bulge sample by the mean SMBH to bulge mass ratio, $\Gamma_{obs} = \langle M_{BH}/M_{bulge}\rangle$. While MM13 chose estimates of bulge masses that excluded contributions to the gravitational potentials from dark matter to estimate the $M_{BH} - M_{bulge}$ relation, we chose to use, where available, bulge-mass estimates including dark matter contributions. This was done in order to obtain the most accurate value of $\Gamma_{obs}$ possible. MM13 were interested in obtaining a self-consistent $M_{BH} - M_{bulge}$ relation, and found that the relation was not significantly affected by the above choice of bulge mass estimates.

To account for the revised sample of SMBH and bulge masses, we scaled the masses of SMBHs in the G11 model by a factor $F$, which is equivalent to adjusting the parameter $f_{BH}$. This equivalence is physically justified for two reasons. First, it is thought that the vast majority of the mass of SMBHs in the local Universe has been built up through accretion in quasar phases (*50*), i.e., the masses of the first generation of black holes are relatively small. In fact, no SMBH seeds are included in the G11 model. Instead, upon the first merger experienced by a pair of galaxies, an SMBH with a mass given by Supplementary Eq. (7) is assumed to be created in the merger remnant. Second, SMBHs are at most a hundredth of the total baryon masses of their host galaxies, indicating that the contribution of SMBHs to the baryonic masses of their host galaxies, and hence the amount of gas accreted in mergers, is largely independent of the SMBH mass. Together, these facts imply that an SMBH at any redshift in the G11 model, having undergone any number of accretion and coalescence episodes with other SMBHs, will have a mass that increases linearly with $f_{BH}$. This was confirmed by examining the SMBH mass functions output by the Croton et al. semi-analytic model (*3*) for different values of $f_{BH}$.

### S6.3 Scaling the masses of the G11 SMBHs

Given that (a) the sample used to measure $\Gamma_{obs}$ comprises only 35 SMBH-galaxy pairs, (b) individual mass measurements show large uncertainty, and (c) the $M_{BH} - M_{bulge}$ relation shows large intrinsic scatter, the value of $F$ has significant uncertainty. To account for this uncertainty when calculating the strength of the GWB, we need to estimate the posterior probability distribution of the factor $F$ given the observed ratio $\Gamma_{obs}$ between SMBH and bulge masses, i.e., $\rho(F \mid \Gamma_{obs})$. This is straightforward to evaluate using Bayes' Theorem:

$$\rho(F \mid \Gamma_{obs}) \propto \rho(\Gamma_{obs} \mid F)\, \rho(F), \quad (8)$$

where $\rho(\Gamma_{obs} \mid F)$ is the probability density of obtaining $\Gamma_{obs}$ for different values of $F$, also referred to as the likelihood of $F$ given $\Gamma_{obs}$. We adopt a uniform prior in $F$ within a reasonable range in $F$, $(0.8 < F < 3.2)$.

We used a Monte Carlo technique to evaluate the posterior distribution $\rho(\Gamma_{obs} \mid F)$. For fixed $F$, we generated $10^5$ random values of $\Gamma$ from the G11 model. Each value was calculated using random selections of 35 SMBH-bulge pairs with the same bulge mass distribution as the sample of MM13. We also generated $10^5$ random values of $\Gamma_{obs}$ using the observational errors. The posterior distribution was then found by estimating the probability density of the distribution of $\log(\Gamma/\Gamma_{obs})$ values at zero. This process was repeated for many values of $F$ in the range $0.8 < F < 3.2$.

Our maximum likelihood estimate of $F$ is 1.9, with the 5th and 95th percentiles of the posterior distribution $\rho(\Gamma_{obs} \mid F)$ lying at $F = 1.46$ and $F = 2.46$ respectively. This is consistent with the updated ratio of the normalization of the $M_{BH} - M_{bulge}$ relation found by MM13.

### S6.4 Predicting $\Omega_{GW}(f)$

We use a technique similar to that outlined in (*9*, hereafter R12) to derive the GW signal from binary SMBHs in the G11 model. R12 found a distribution of the number of observable binary SMBHs per unit GW frequency per unit frequency-independent GW power, $h_0$, from the G11 model, and fitted this distribution with an analytic function. The GW power is defined as

$$h_0 = \sqrt{\frac{32}{5}} \frac{(GM_C)^{5/3}}{c^4 D(z)} (\pi(1+z))^{2/3}, \quad (9)$$

where $M_C$ is the chirp mass, $c$ is the vacuum speed of light, $G$ is Newton's gravitational constant, and $D(z)$ is the comoving coordinate distance at redshift $z$. The distribution, $\Phi$, is defined to be

$$\Phi(h_0, z, f) = 4\pi \frac{dN}{dh_0} \frac{d^2 V_c}{d\Omega dz} \frac{dz}{dt} \frac{dt}{df}, \quad (10)$$

where $N$ is the number of binary SMBHs per unit comoving volume per unit solid angle on the sky. While the derivative $dz/dt$ is straightforward to evaluate from cosmological theory, the derivative $dt/df$ depends on the binary chirp mass, redshift and frequency (see Eq. 15 of R12). We therefore cannot evaluate $\Phi$ at values of $F \neq 1$ from the fitted form alone.

As in R12, we first used the G11 model to find the numbers of observable SMBH-SMBH coalescence events in different bins of $h_0$ and $z$, i.e.,

$$N(h_0, z) \approx 4\pi \frac{dN}{dh_0} \frac{d^2 V_c}{d\Omega dz} \Delta h_0 \Delta z, \quad (11)$$

where $\Delta h_0$ and $\Delta z$ are the bin-widths. As outlined in R12, the G11 model does not predict a unique $N(h_0, z)$; we instead formed 1000 realisations, and summed them to get $N_{1000}(h_0, z)$. In order to well-approximate the mean form of $N(h_0, z)$, we fitted a broken power-law function to the distribution at each redshift:

$$N_{\text{fit},1000}(h_0, z) = n \, (h_0/p_h)^\beta \, (1 + h_0/p_h)^\gamma, \quad (12)$$

where $n$, $p_h$, $\beta$, and $\gamma$ are free parameters. This modeling is essential when calculating the average properties of the GW signal.

In order to appropriately account for $h_0$ bins with no events, prevalent at the high-$h_0$ end of the distributions at all redshifts, we re-fit the distributions assuming Poisson-distributed counts in each bin above the break in the power law, $p_h$. Maximum-likelihood fits were performed to single power laws for $h_0$ values above $p_h$ using Markov Chain Monte Carlo sampling of the likelihoods. We found $\Phi$ by combining Supplementary Eqs. (10), (11), and (12) (dividing each $N_{\text{fit},1000}(h_0, z)$ by 1000), and then calculated $\Omega_{GW}(f)$ by evaluating

$$\Omega_{GW}(f) = \frac{2\pi^2}{3H_0^2} \int_{h_{0,\min}(z)}^{h_{0,\max}(z)} dh_0 \int_0^\infty dz \, \Phi(h_0, z, f) f^{2/3}. \quad (13)$$

The lower bound of the $h_0$ integral at each redshift, $h_{0,\min}(z)$, was set to correspond to the value of $h_0$ corresponding to a binary with chirp mass $10^6 \, M_\odot$ at $z = 6$. The upper bound, $h_{0,\max}(z)$, was set to the larger of either the value of $h_0$ of the strongest source in $N_{1000}(h_0, z)$ or the value above which one source was expected in the given redshift bin. The probability distribution $\rho(F \mid \Gamma_{\text{obs}})$ into a probability distribution $\rho_M(\Omega_{GW})$, depicted in Fig. 2 of the main text, using Supplementary Eq. (13), and by calculating the distribution $\Phi$ for different values of $F$.

For each value of $F$, new distributions $N_{\text{fit},1000}(h_0, z)$ were evaluated. We assumed that $\Omega_{\text{GW}}(f) \propto f^{2/3}$, which is expected for circular binaries evolving under GW emission alone. We do not need to account for binaries that have reached the last stable orbits, because binaries in the mass and mass ratio ranges under consideration are expected to reach their last stable orbits at frequencies much larger than $f_{\text{PPTA}}$.

The use of Poisson fitting to the source distributions at each redshift, the method of setting the $h_0$ bounds, and the removal of a factor of $(4/3)^{1/2}$ from the definition of the mean strain amplitude of a binary together represent the major differences between the methods used here and in R12 to calculate the GW signal predicted by the G11 model. If instead of using Poisson fitting, standard least-squares fitting is used, the fitted functions are biased low above $p_h$ at every redshift. In R12, an analytic function was fitted using least-squares methods to the redshift-integrated distribution of binary SMBHs in $h_0$. The upper bound, $h_{0,\text{max}}(z)$, was set at to be the value above which the averaged distribution was non-continuous.

For $F = 1$, we find $A = 8.0 \times 10^{-16}$. For this work, we therefore simulate ToA variations induced by the GWB with non-Gaussian statistics for $F \neq 1$ by scaling the strengths of individual GW sources simulated with $F = 1$ to produce the expected mean $\Omega_{\text{GW}}(f)$ corresponding to a given $F$. We verified this method by generating multiple realizations of lists of binaries from $\Phi(h_0, z, f)$ for $F = 1$, $2$ and $3$, and comparing the power spectra of the resulting ToA variations with those generated using the simpler technique using Anderson-Darling tests.

### S6.5: Extending the G11 model of the GWB

The G11 model is designed to produce SMBHs at $z = 0$ with masses that are, on average, a fixed fraction of the masses of their host bulges regardless of the type of galaxy they reside in. This is consistent with a number of observational results (*1, 2, 48*). However, it is possible that galaxies that are well modeled by Sérsic luminosity profiles, and galaxies with partially depleted cores, follow different $M_{\text{BH}} - M_{\text{bulge}}$ relations (*51*). *Core* galaxies tend to have greater masses than *Sérsic* galaxies, and are thought to have formed through dissipationless dry mergers, whereas Sérsic galaxies are thought to have formed through dissipational, secular processes. The $M_{\text{BH}} - M_{\text{bulge}}$ relation for core galaxies is linear, with a mean ratio $\Gamma$ that is 0.1 dex greater than that fitted by MM12 for their entire sample (*51*). While attempting to tune the G11 model to match the proposed broken power-law $M_{\text{BH}} - M_{\text{bulge}}$ relation would result in a higher GWB amplitude, this would not reflect the fact that the model does not account for the multiple dominant modes of SMBH growth that would be required to reproduce these observations.

### S6.6: An alternative physical model for the GWB

The alternative physical model that we consider (*29*) uses an adaptive mesh refinement code to perform cosmological hydrodynamic simulations of a cluster and a void environment, which trace the assembly of galaxies and SMBHs at redshifts $z < 4$. This model differs from the Millennium based model presented in S6.1 to S6.4 in the following ways:

- This model numerically traces the evolution of the baryonic components in the simulations along with the dark matter, rather than using a semi-analytic framework. The prescriptions for various physical processes are also slightly different; see (*29*) and (*4*) and references therein for details.

- The assumed cosmological parameters in this model are from the 7 yr Wilkinson Microwave Anisotropy Probe results, rather than from the 1 yr results.

The resulting prediction for $\Omega_{GW}(f_{PPTA})$ is higher than the prediction based on the G11 model.

## S7: Comparing the GWB predictions to the PPTA data

Here, we describe how we calculate the probability, $\Pr(M)$, of a given model prediction for $\Omega_{GW}$ (denoted $M$) being consistent with the PPTA constraints on $\Omega_{GW}$. The model prediction is represented by the probability density function $\rho_M(\Omega_{GW})$:

$$\rho_M(\Omega_{GW}) = \frac{d\Pr(M|\Omega_{GW})}{d\log\Omega_{GW}}. \quad (14)$$

This can be interpreted as the conditional probability density of the model being true given a value of $\Omega_{GW}$.

By conducting simulations with different injected GWB strengths $A_{sim}$ as described in Section S2, we empirically evaluate the probabilities of different values of $A_{sim}$, and hence different values of $\Omega_{GW}$ (i.e., $\Pr(\Omega_{GW})$). For example, for $\Omega_{GW}$ corresponding to $A_{95}$, $\Pr(\Omega_{GW}) = 0.05$. We plot $\rho_M(\Omega_{GW})$ for each model under consideration, and $\Pr(\Omega_{GW})$ for both Gaussian and non-Gaussian GWBs in Fig. 2 of the main text.

From the law of total probability,

$$\Pr(M) = \int_{-\infty}^{\infty} d\log\Omega_{GW} \rho_M(\Omega_{GW}) \Pr(\Omega_{GW}). \quad (15)$$

For both a Gaussian and non-Gaussian GWB, we find that $\Pr(\Omega_{GW})$ is well modeled using a complementary Gaussian error function. For a Gaussian GWB, we have

$$\Pr(\Omega_{GW}) = \int_{\log\Omega_{GW}}^{\infty} d\log\Omega'_{GW} \frac{1}{\sqrt{2\pi(0.25)^2}} \exp\left[-\frac{(\log\Omega'_{GW}+9.37)^2}{2(0.25)^2}\right], \quad (16)$$

and for a non-Gaussian GWB,

$$\Pr(\Omega_{GW}) = \int_{\log\Omega_{GW}}^{\infty} d\log\Omega'_{GW} \frac{1}{\sqrt{2\pi(0.33)^2}} \exp\left[-\frac{(\log\Omega'_{GW}+9.34)^2}{2(0.33)^2}\right], \quad (17)$$

For the G11 model, we find that $\rho_M(\Omega_{GW})$ can be modeled to be the sum of two Gaussian functions:

$$\rho_M(\Omega_{GW}) = \frac{0.983}{\sqrt{2\pi(0.115)^2}} \exp\left[\frac{-(\log\Omega_{GW}+9.35)^2}{2(0.115)^2}\right] + \frac{0.017}{\sqrt{2\pi(0.123)^2}} \exp\left[\frac{-(\log\Omega_{GW}+9.54)^2}{2(0.123)^2}\right], \quad (18)$$

In addition to the Millennium-based model, we examine how the PPTA limit on the strength of the GWB can be used to constrain two recently developed predictions for the strength of the GWB.

For the *empirical synthesis* of models (26), we take for $\rho_M(\Omega_{GW})$ the prediction given the most recent SMBH and bulge mass measurements. These results yield a lognormal distribution for the probability density function of

$$\rho_M(\Omega_{GW}) = \frac{1}{\sqrt{2\pi(0.610)^2}} \exp\left[\frac{-(\log\Omega_{GW} + 9.44)^2}{2(0.610)^2}\right]. \quad (19)$$

For the *merger-driven model* (*25*), the probability distribution is

$$\rho_M(\Omega_{GW}) = \frac{1}{\sqrt{2\pi(0.571)^2}} \exp\left[\frac{-(\log\Omega_{GW} + 8.53)^2}{2(0.571)^2}\right]. \quad (20)$$

For the *alternative physical model* (*29*), the probability distribution is

$$\rho_M(\Omega_{GW}) = \frac{1}{\sqrt{2\pi(0.260)^2}} \exp\left[\frac{-(\log\Omega_{GW} + 9.27)^2}{2(0.260)^2}\right]. \quad (21)$$

By substituting expressions for $\Pr(\Omega_{GW})$ and $\rho_M(\Omega_{GW})$ for the different models into Supplementary Eq. (15), we find $\Pr(M)$.

Supplementary Figure

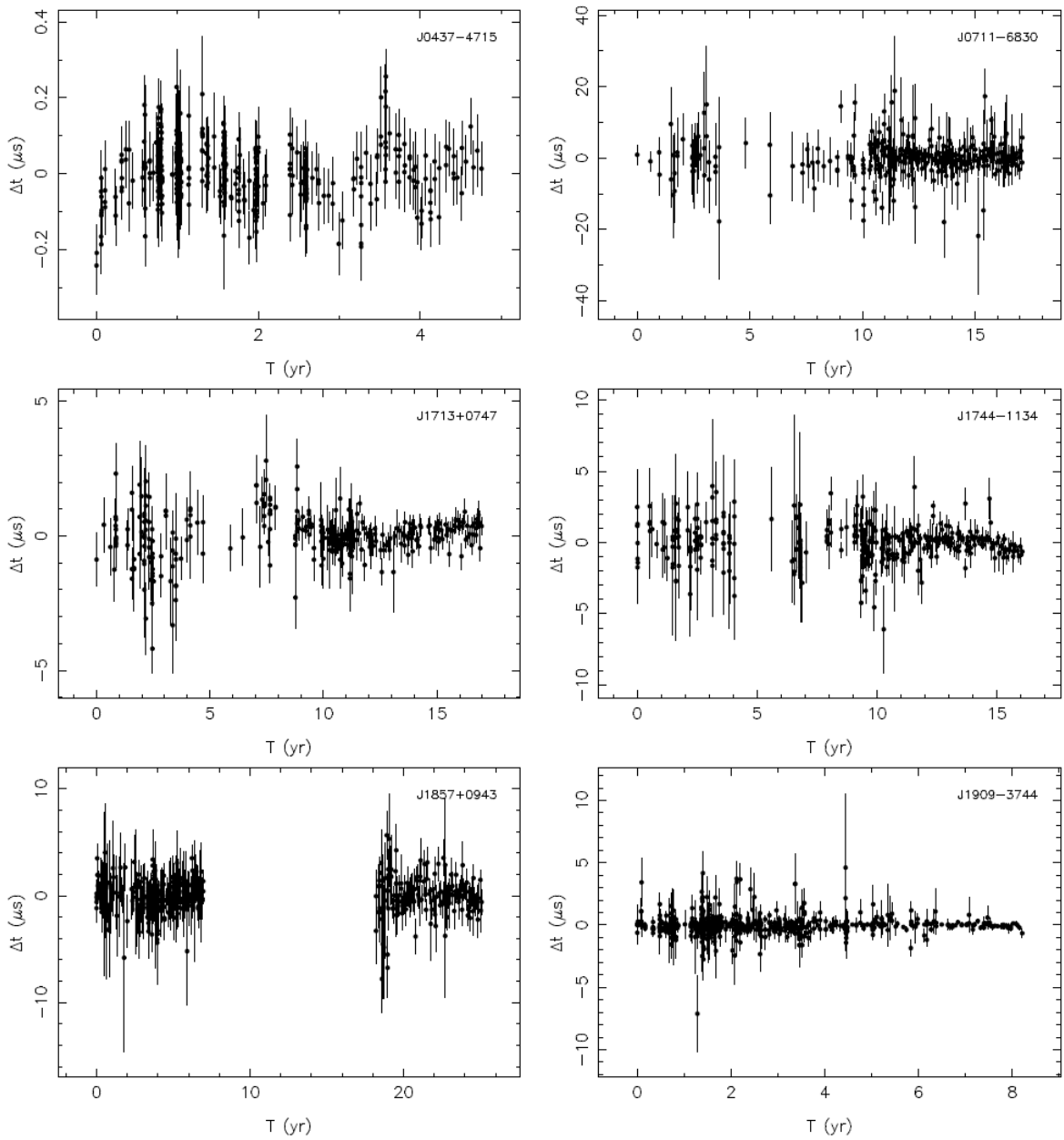

**Supplementary Fig. S1:** Post-fit timing residuals for the PPTA pulsars (*14*) used to place a limit on the GWB amplitude.

Supplementary Tables

| PSR | EQUAD (μs) | $\sigma_{ToA}$ (μs) | $T_{span}$ (yr) | $\lambda_{best}$ (cm) |
|---|---|---|---|---|
| J0437-4715 | 0.065 | 0.066 | 4.8 | 10 |
| J0711-6830 | 1.5 | 2.6 | 17.1 | 20 |
| J1713+0747 | 0.25 | 0.51 | 17.0 | 10 |
| J1744-1134 | 0.50 | 0.73 | 16.1 | 20 |
| J1857+0943 | 0.65 | 1.16 | 25.1 | 20 |
| J1909-3744 | 0.17 | 0.24 | 8.2 | 10 |

**Table S1:** Data sets used in timing analysis. We list the quadrature errors added to the ToA uncertainties, EQUAD; the weighted root mean square of the post-fit ToAs, $\sigma_{ToA}$; the total observing span, $T_{span}$; and the best band wavelength $\lambda_{best}$ in the DR1 data set.

| Data Set | A (x $10^{-15}$) | $A_{95}$ (x $10^{-15}$) |
|---|---|---|
| Open, 1 | 50 | 52 |
| Open, 2 | 50 | 59 |
| Open, 3 | 10 | 13 |
| Closed, 1 | 10 | 12 |
| Closed, 2 | 60 | 70 |
| Closed, 3 | 5 | 7.2 |

**Table S2:** Limits on the GWB strength from the IPTA data challenge data sets. The IPTA data challenge contained six mock data sets. Details of the data sets can be found on the IPTA website (http://www.ipta4gw.org).